\documentclass[aps, prd, reprint]{revtex4-1}
\usepackage[utf8]{inputenc}
\usepackage{lipsum}
\usepackage{graphicx}
\usepackage[T1]{fontenc}
\usepackage{url}
\usepackage{bm}
\usepackage{epstopdf}
\usepackage{amstext} 
\usepackage{amsmath, bm}
\usepackage{fixmath} 
\usepackage{inputenc}
\usepackage{physics}
\usepackage[singlelinecheck=true,justification=raggedright]{caption}
\usepackage{subcaption}
\usepackage{threeparttable}
\usepackage{float}
\usepackage{svg}
\begin{document}

\title{A torsion-balance search for ultra low-mass bosonic dark matter}

\author{E.\,A.~Shaw}
\author{M.\,P.~Ross}
\author{C.\,A.~Hagedorn}
\author{E.\,G.~Adelberger}
\author{J.\,H.~Gundlach}
\affiliation{Center for Experimental Nuclear Physics and Astrophysics, Box 354290, University of Washington, Seattle, Washington 98195-4290} 

\begin{abstract}
We used a stationary torsion balance with a beryllium-aluminum composition dipole to search for ultra low-mass bosonic dark matter coupled to baryon minus lepton number. We set 95\% confidence limits on the coupling constant $g_{\rm B-L}$ for bosons with masses between $10^{-18}$ and $10^{-16}$~eV/$c^2$ with the best performance at $m_{\rm DM} = 8\times 10^{-18}$ eV/$c^2$ constraining $g_{B-L}(\hbar c)^{-1/2} < 1 \times 10^{-25}$. This provides a complimentary limit to equivalence-principle experiments that search for ultra low-mass bosons as force-mediating particles.
\end{abstract}
\pacs{95.35+d,98.35Gi,14.80.Va}
\maketitle

\section{Introduction}
Despite an abundance of indirect astrophysical evidence for the existence of cold dark matter \cite{be:10}, the nature of this phenomenon remains a mystery. Many direct detection searches have assumed that the dark matter consists of supersymmetry-inspired fermions called WIMPs. However, despite considerable effort, no evidence has been found for supersymmetry \cite{supersymmetry} nor for WIMPs \cite{wimps}. Focus has now shifted to dark matter consisting of ultra low-mass bosons (ULMBs). It has been shown that scalar, pseudovector or vector bosons with a wide range of masses between 10$^{-22}$ and 100~eV/$c^2$ are viable dark matter candidates \cite{Hui}. 
For dark matter masses $m_{\rm DM} \le 1$~eV/$c^2$, the dark matter number density is large enough to make the dark matter behave as a classical field. Dark matter particles bound in our galaxy must be highly non-relativistic with a local velocity distribution peaked at $v_{\rm DM} \approx 10^{-3}c$ \cite{Bovy_2012}. The field would oscillate at a frequency 
\begin{equation}
f_{\rm DM}=\frac{m_{\rm DM} c^2}{h} [1+(v_{\rm DM}/c)^2/2],\label{eq:1}
\end{equation}
where $c$ is the speed of light and $h$ is Planck's constant. This corresponds to a central Compton frequency $f_{\rm DM}=m_{\rm DM} c^2/h$ with a fractional spread $\delta f_{\rm DM}/f_{\rm DM}=(v_{\rm DM}/c)^2/2 \approx 10^{-6}$. This field should be coherent for times less than the phase coherence time
\begin{equation}
t_{\rm c}=1/\delta f_{\rm DM}~,\label{eq:2}
\end{equation}
{\em i.e.} about $10^6$ oscillation periods.


Many searches for these dark-matter waves have been recently reported. Ultra-stable clocks \cite{dilaton, atomicclocksdilaton} and gravitational-wave detectors \cite{GWdilaton} constrain the couplings of scalar dilaton ULMBs with masses between $10^{-21}$ and $10^{-5}$ eV/$c^2$. A wide variety of experimental approaches constrain the couplings of pseudo-scalar axion or axion-like ULMBs. Spin-precession experiments \cite{nEDM, SPIN, Smorra2019} constrain the coupling of pseudoscalar axion-like ULMBs with masses between $10^{-23}$ and $10^{-18}$ eV/$c^2$. Axion-to-photon conversion in ultra-sensitive electromagnetic cavities immersed in strong magnetic fields \cite{Sikivie} are searching for the Peccei-Quinn axion \cite{peccei} in the mass range $10^{-6}$ eV/$c^2$ $\lesssim m_{DM}\lesssim 10^{-3}$ eV/$c^2$ \cite{HAYSTAC2021, CAST}. The ADMX collaboration \cite{ADMX2021} has definitively excluded such dark-matter axions with masses between 2.7 $\mu$eV/c$^2$ and 4.2 $\mu$eV/c$^2$.
%
Fewer searches for vector ULMBs have been reported, and these focus on dark photons \cite{ADMXhiddenphoton, superQubit}.
Here we present limits on vector ULMBs coupled to B-L (B and L are baryon number and lepton number, respectively); this coupling is particularly interesting because B-L is conserved in many unified theories. A coherent vector boson wave couples to laboratory objects like a time-varying "B-L electric" field $\tilde{\bm{E}}$ interacting with "B-L charges" $\tilde{q}$ on the objects with a coupling constant $\tilde{g}$. 
If the dark matter consists predominantly of such bosons, $\tilde{E} \approx \sqrt{\rho_{DM}\hbar c} = 7.7\times 10^{3}$ eV/m \cite{gr:13} where $\rho_{DM} \approx 0.3$ GeV/cm$^3$ is the local dark-matter density. The direction of $\tilde{\bm{E}}$ is expected to be random but steady for times less than $t_{\rm c}$ \cite{gr:13,nelson}. The force on a "charge" $\tilde{q}$ is
\begin{equation}
\bm{F}_{DM} = \tilde{q}\tilde{g}(\hbar c)^{-1/2} \tilde{\bm{E}}.
\end{equation}
The B-L "charge" of a test-body consisting of electrically neutral atoms is $(q_{\rm B-L}/\mu)(m_T/u) = (N/\mu)(m_T/u)$, where $N$ is the number of neutrons, $m_T$ is the test-body mass, and $\mu$ is atomic mass in atomic mass units (u). We detect these forces using a torsion pendulum containing a "charge" dipole. The differential acceleration (force/mass) of the 2 "charges", 
\begin{equation}
\Delta a_{B-L} = g_{B-L} \sqrt{\rho_{DM}}\, \Delta_{B-L}/u~,
\end{equation}
applies a torque on the pendulum that is our ULMB signal
($\Delta_{B-L} = (N_1/\mu_1)-(N_2/\mu_2)$ is the difference in the charge-to-mass ratios of the two atomic species in the dipole). 

\begin{figure}[h!]
\centering \includegraphics[width=0.5\textwidth]{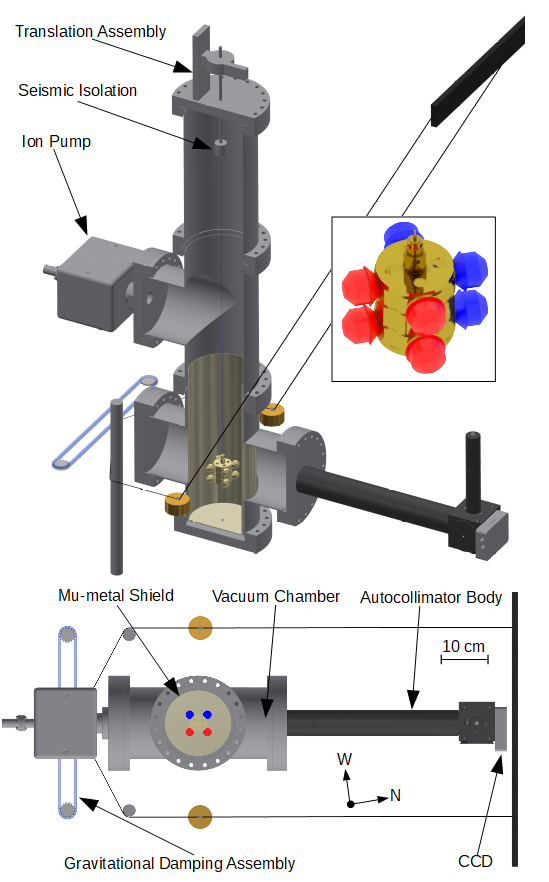}
\caption{Perspective and top views of our apparatus. The insert shows the 8 test-mass pendulum \cite{CQG2012} (red and blue indicate the composition dipole of four Be and four Al test masses) that hangs within a mu-metal shield in the vacuum chamber. A motor-driven roller chain is on the left side of the torsion balance. A cord attached to the roller chain moves the two 486 g brass cylinders in opposite directions effectively rotating a quadrupole field with respect to the balance. The object protruding on the right is the autocllimator that measures the pendulum twist. The N and W vectors indicate the orientation of the instrument and define the lab coordinate system: $\hat{\bm{x}}$ points North, $\hat{\bm{y}}$ points West, and $\hat{\bm{z}}$ is local vertical.}
\label{Apparatus}
\end{figure}

\section{Experimental Setup}
\begin{figure}[h!]
\centering \includegraphics[width=\columnwidth]{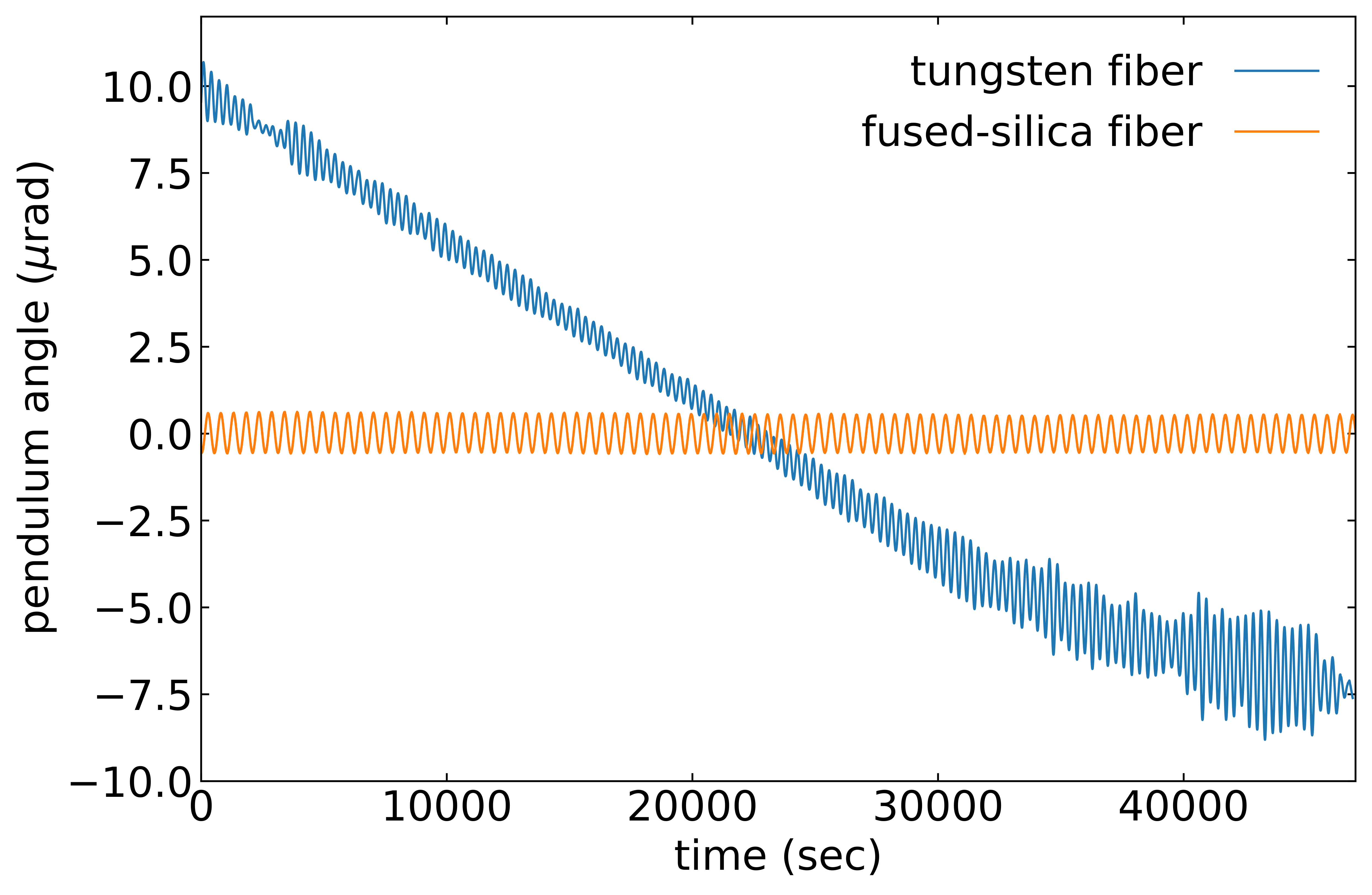}
    \caption{Comparison of fused-silica (orange) and tungsten (blue) fibers. The lower drifts in fused silica ($<1$~$\mu$rad/day) are much smaller than tungsten ($>10$~$\mu$rad/day). The rapid oscillations are the resonant free oscillations of the torsion oscillator; these are driven by thermal effects and environmental vibrations. The higher Q of fused silica is obvious.}
\label{fig:fiber}
\end{figure}

\begin{figure}[h!]
\centering \includegraphics[width=\columnwidth]{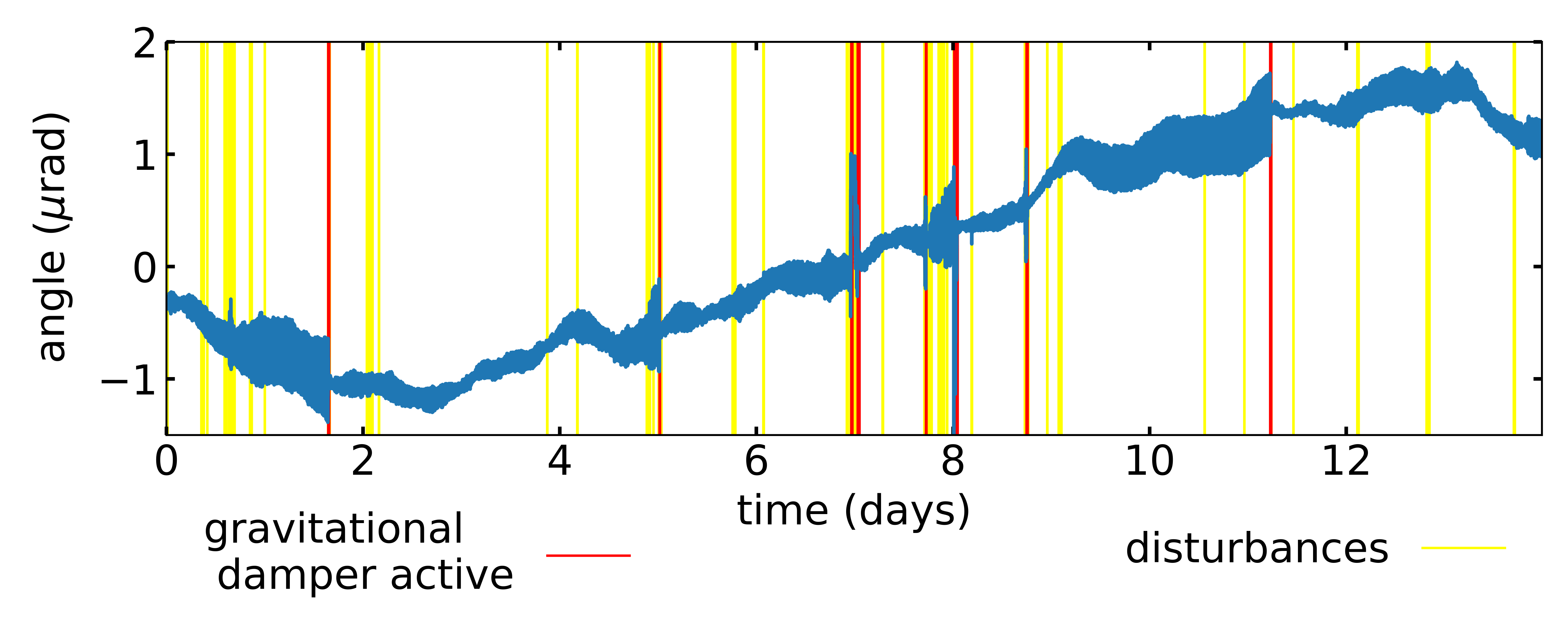} \includegraphics[width=\columnwidth]{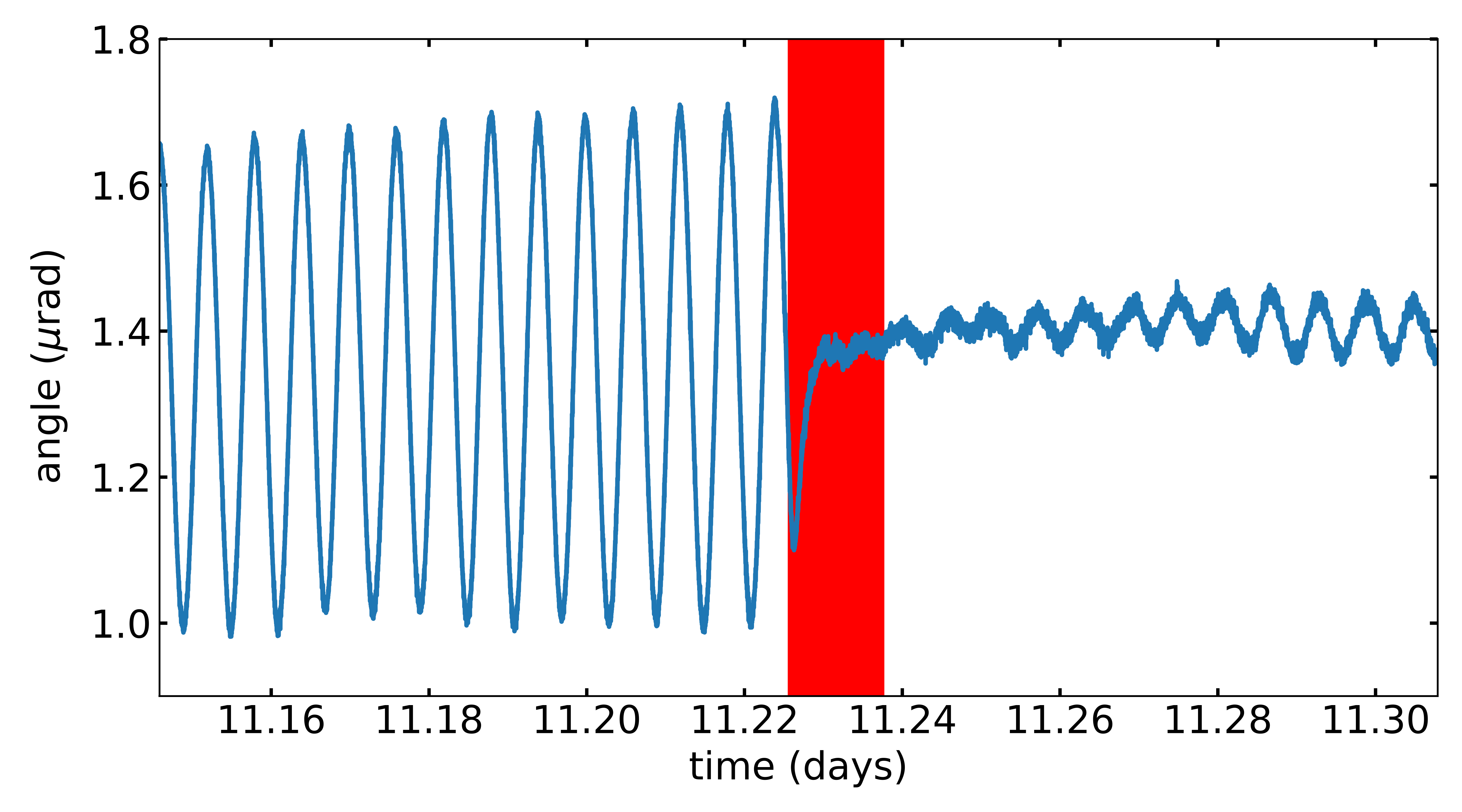} \includegraphics[width=\columnwidth]{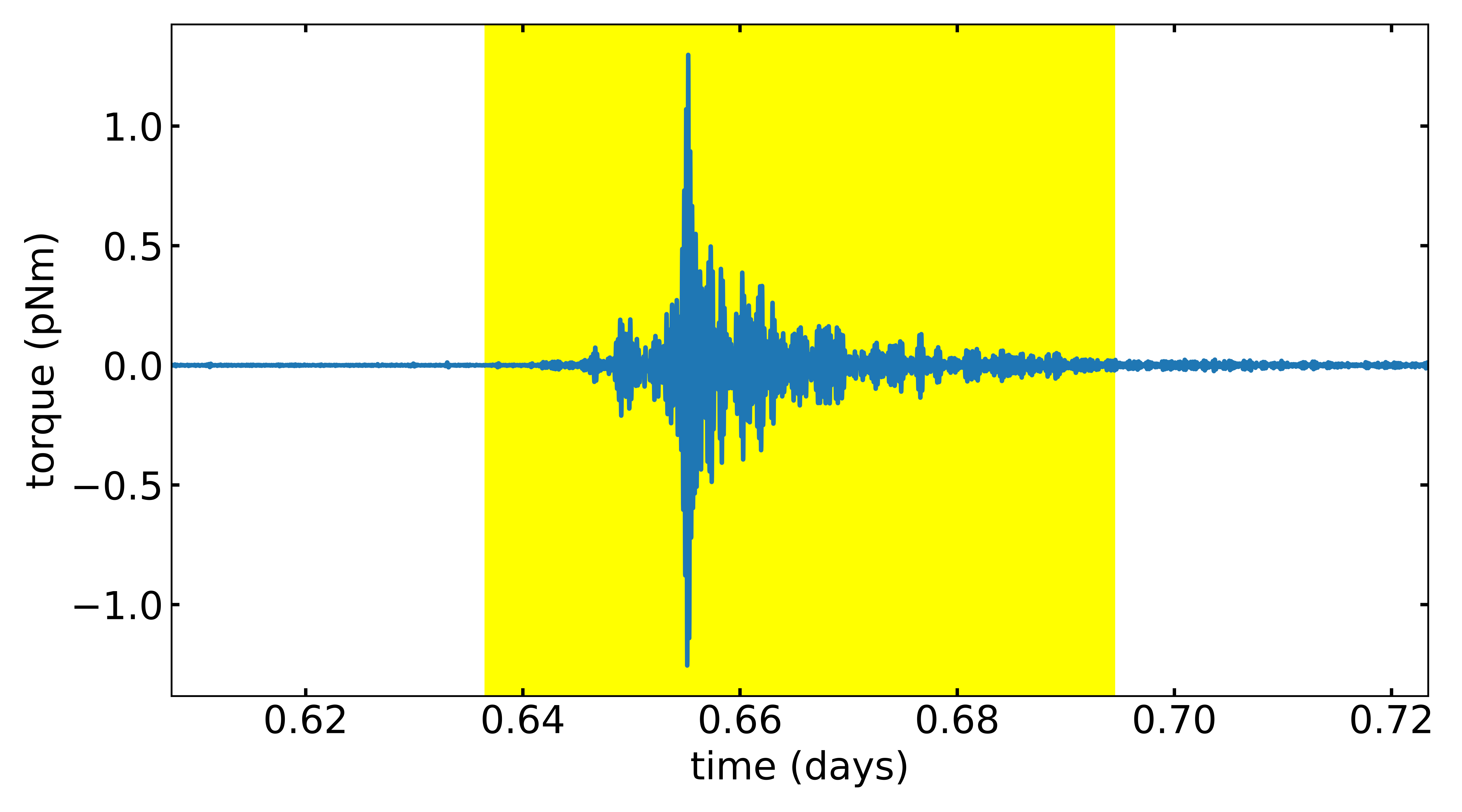} \caption{\textbf{Top:} Two weeks of data from our ULMB campaign. Red bars indicate times when the gravitational damper was active, while yellow bars indicate seismic events. These data were excluded. The gravitational damper allowed us to maintain a small free-oscillation amplitude. \textbf{Middle:} Gravitational damper performance on day 11 in top plot. The damper efficiently removed residual oscillation. \textbf{Bottom:} Example of filtered torque data dropped to exclude an earthquake (on day 1 where there is a noticeable spike in the angle timeseries).} \label{fig:EPBUDMTimeSeries}
\end{figure}

We searched for the frequency-dependent signatures of B-L coupled ULMBs using a stationary torsion-balance with a pendulum consisting of a Be-Al composition dipole for which $\Delta_{B-L} = 0.0359$. Fig. \ref{Apparatus} shows an overview of our apparatus. The pendulum was suspended by a fused-silica fiber manufactured in our lab. The observed resonant frequency $f_0 = (2\pi)^{-1}\sqrt{\kappa/I} = 1.934$ mHz and the rotational inertia $I = 3.78\times 10^{-5}$ kg~m$^2$ (obtained from a detailed model of the pendulum) yielded a torsion constant of $\kappa = 5.58$ nNm/rad. We observed a quality factor $Q =$ 460,000 corresponding to a thermal noise \cite{saulson},
\begin{equation}
	\tau(f) = \sqrt{\frac{2k_B T\kappa}{\pi Q f}},
\end{equation}
which has a value at 1 mHz of $1.8\times 10^{-16}$ Nm/$\sqrt{\rm Hz}$. The corresponding value of our tungsten fiber ($\kappa = 2.4$ nJ/rad and $Q = 6700$ \citep{CQG2012}) is $9.6\times 10^{-16}$ Nm/$\sqrt{\rm Hz}$. Furthermore, the low-frequency drift and thermal-susceptibility of silica is smaller by at least an order of magnitude (Fig.~\ref{fig:fiber}). The pendulum sits within a vacuum chamber and is surrounded by a mu-metal shield. We measured the pendulum angle, $\theta$, using a multi-slit autocollimator \cite{autocollimator}. 
We kept $\theta$ within the autocollimator's linear range using a novel gravitational damper, consisting of an adjustable mass quadrupole that applied a controlled torque on the pendulum. The performance of the damper is displayed in Fig. \ref{fig:EPBUDMTimeSeries}. 

\section{Data Analysis}
Our data set consisted of 91.4 days of angle measurements taken at a sampling interval $\Delta = 1.00212$ s over a span of $S =  114$ days. We searched for signals from dark matter with 491,019 different possible masses between $0.4\times 10^{-18}$ and $206.8 \times 10^{-18}$ eV/$c^2$ ($f_{\rm DM}$ from 0.1 mHz to 50 mHz) in steps of $4 \times 10^{-22}$ eV/$c^2$ ($1/S = 101.6$ nHz). At each assumed value of $m_{\rm DM}$ we obtained the corresponding torque on the pendulum 
\begin{align}
\tau(t) =& I\ddot{\theta}(t)+\kappa\left(1+\frac{i}{Q}\right)\theta(t) \nonumber \\
 \approx& I\ddot{\theta}(t)+\frac{\kappa}{2\pi f Q}\dot{\theta}(t)+\kappa\theta(t).
\label{torque_eqn}
\end{align}
The first derivative term gives the exact response for a single-frequency signal. Hence, we used numerical derivatives \cite{derivatives} to estimate the torque as
\begin{align}
\tau_j =& \frac{I}{\Delta^2}\left(-\frac{\theta_{j-2}}{12}+\frac{4\theta_{j-1}}{3}-\frac{5\theta_j}{2}+\frac{4\theta_{j+1}}{3}-\frac{\theta_{j+2}}{12}\right) \nonumber \\
 &+\frac{\kappa}{2\pi f_{DM} Q\Delta}\left(\frac{\theta_{j-2}}{12}-\frac{2\theta_{j-1}}{3}+\frac{2\theta_{j+1}}{3}-\frac{\theta_{j+2}}{12}\right) \nonumber \\
 &+\kappa\theta_j \label{torque_cor},
\end{align}
where $\theta_j$ and $\tau_j$ are the angle and torque at the j$^{\mathrm{th}}$ measurement. 

We applied a lowpass FIR filter \cite{hamming} to avoid high-frequency spectral leakage dominated by autocollimator noise. We divided the torque data into sidereal-day length cuts and removed data points when the gravitational damper was active or during noticeable glitches in the data. We adjusted the starting time of each cut so that fewer than 10\% of the points were removed and no damping event occurred during the cut. These data quality measures, which gave us a total of 40 cuts, ensured acceptable covariance between the basis functions (see Appendix). From here the analysis was similar to that introduced in Ref.~\cite{SPIN} and summarized below.

The dark matter torque on our dipole is $\bm{\tau}_{\rm DM}=\Delta q_{tot} g_{\rm B-L}(\hbar c)^{-1/2}(\bm{d}\times\tilde{\bm{E}}_{lab})/2$ where $\Delta q_{tot}$ is the difference of the total B-L charge on the dipole and $\bm{d} = 3.77\,(\cos{\gamma_d}\,\hat{\bm{x}}-\sin{\gamma_d}\,\hat{\bm{y}})$ cm is the vector from one element of the composition dipole to the other with azimuthal angle $\gamma_d=-80.7\pm 5$. But our instrument is only sensitive to the vertical component so that
\begin{equation} 
\tau_{\rm DM} = \frac{m_d g_{\rm B-L}\Delta_{B-L}}{2u\sqrt{\hbar c}}\bm{p}\cdot\tilde{\bm{E}}_{lab}\ .
\end{equation} 
Here $m_d=19.4$ g is the mass of four of our test bodies and $\bm{p}=\hat{\bm{z}}\times\bm{d}$. 

We constrained ULMB waves polarized along arbitrary directions in geocentric celestial coordinates $\hat{\bm{X}}$, $\hat{\bm{Y}}$, $\hat{\bm{Z}}$ (the origin is the center of the Earth, $\hat{\bm{X}}$ points towards the sun at the vernal equinox, $\hat{\bm{Z}}$ points North and passes through the center of the Earth, and $\hat{\bm{Y}} = \hat{\bm{Z}}\times\hat{\bm{X}}$). The anticipated signals are coherent for $\approx 10^6$ oscillations (see equation \eqref{eq:2}), which for the highest fitted frequency of $50$ mHz corresponds to 231 days, 2 times longer than the span of our data. 

Following the procedure introduced in Ref.~\citep{SPIN} we generated 6 science basis functions that evaluated $\bm{p}\cdot\tilde{\bm{E}}_{\rm lab}$ for unit-strength fields along the $\hat{\bm{X}}$, $\hat{\bm{Y}}$, and $\hat{\bm{Z}}$ directions ($K_{\rm Xp}$, $K_{\rm Yp}$, and $K_{\rm Zp}$) with each having a cosine and sine phase. We fit our torque data with
\begin{align}
\tau_j =& K_{\mathrm{Xp}}(t_j)[a_{\mathrm{Xcos}}\cos{(\omega_{\rm DM}t_j)}+a_{\mathrm{Xsin}}\sin{(\omega_{\rm DM}t_j)}] \nonumber \\
&+ \left(\mathrm{similar\, terms\, for\, Y\, and\, Z}\right) \nonumber \\
&+ \left(\mathrm{instrumental\, parameters}\right) \label{torque_basis}
\end{align}
where $\omega_{\rm DM} = m_{\rm DM}c^2/\hbar$. Note that this expression is perfectly general as any phase is a simple linear combination of sine and cosine amplitudes. We computed the components of $\hat{\bm{X}}$, $\hat{\bm{Y}}$, and $\hat{\bm{Z}}$ parallel to $\hat{\bm{p}}$ using the \texttt{AstroPy} library \cite{astropy} to find the altitude and azimuth ($\alpha$ and $\gamma$) of these unit vectors in the lab frame. For example, the projection of $\hat{\bm{X}}$ onto $\hat{\bm{p}}$ is
\begin{equation}
K_{\mathrm{Xp}}(t_j) = -\cos\alpha_{\mathrm{X}}^j\sin(\gamma_{\mathrm{X}}^j-\gamma_{\mathrm{d}})\ .
\end{equation}
Here $\alpha_{\mathrm{X}}^j$ and $\gamma_{\mathrm{X}}^j$ are the altitude and azimuth of $\hat{X}$ for the j$^{\rm th}$ measurement. We included 6 additional instrumental parameters to account for the behavior of the equilibrium angle: 2 for offset and linear drift ($a_0$ and $a_1$) and 4 for daily variations with 24 and 12 hour periods ($a_{\rm cos24}$, $a_{\rm sin24}$, $a_{\rm cos12}$, and $a_{\rm sin12}$). These allowed for daily temperature and tilt variations. The analysis in the Appendix shows that for Compton frequencies of interest (greater than 10x the sidereal frequency) there is negligible covariance between the science and instrumental basis functions.

\begin{figure}[h!]
\centering \includegraphics[width=\columnwidth]{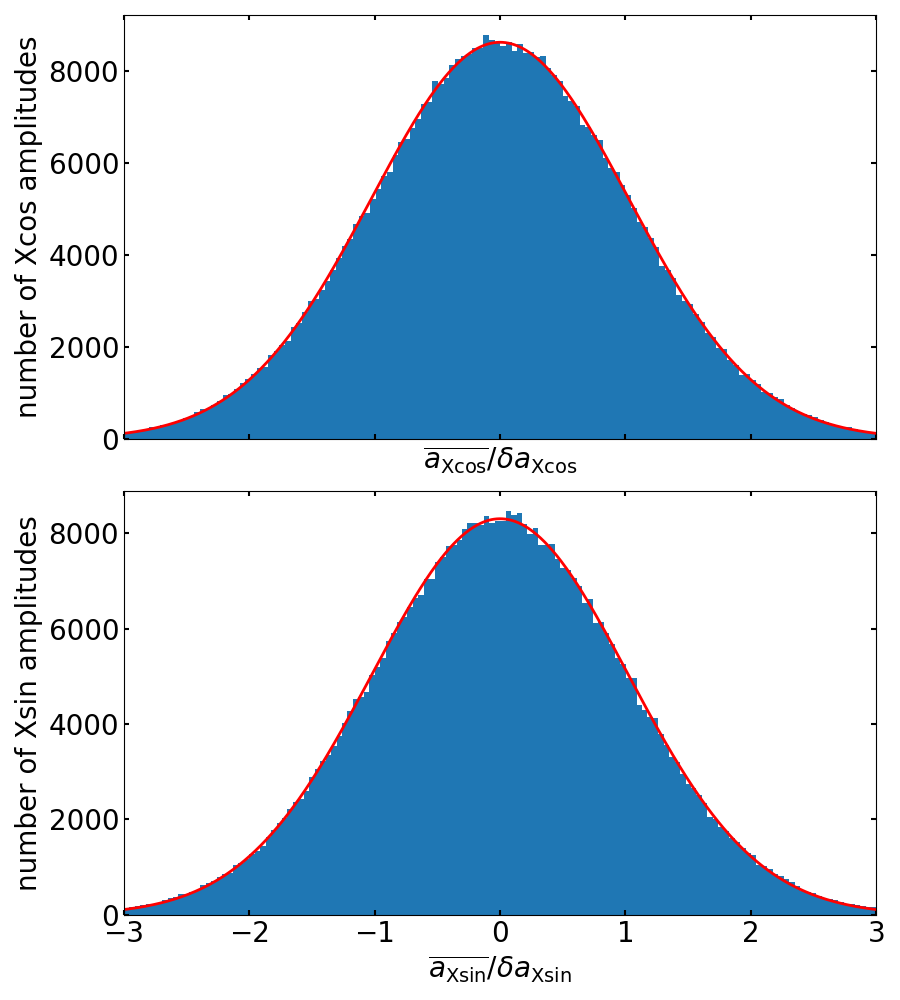} \includegraphics[width=\columnwidth]{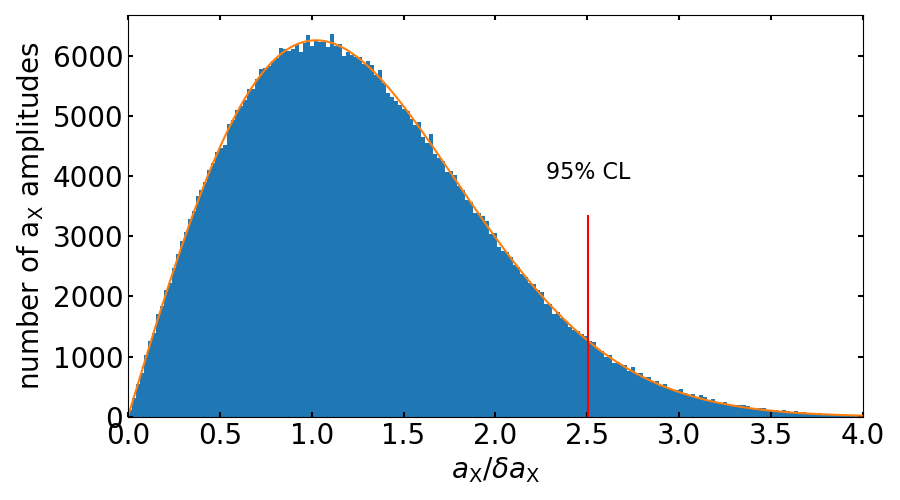}
\caption{\textbf{Upper 2 panels}: histograms of $\overline{a_{\mathrm{Xcos}}}/\delta a_{\mathrm{Xcos}}$ and $\overline{a_{\mathrm{Xsin}}}/\delta a_{\mathrm{Xsin}}$ for all 
491,019 masses. Results are characterized by zero-mean Gaussians with $\sigma = 1.025(1)$ and $1.022(1)$ (shown by the red curves). Similar results are observed for Y and Z. \textbf{Lower panel}: histogram of corresponding $a_{\mathrm{X}}/\delta a_{\mathrm{X}}$ values. As expected these follow a Rayleigh distribution with scale parameter $\sigma = 1.023(1)$ (shown by red curve). The results for Y and Z had scale parameters $1.023(1)$ and $1.021(1)$, respectively.}
\label{fig:GaussRayleighStats}
\end{figure}

For every assumed mass, we extracted the science parameters from linear least-squares fits of the 40 cuts. For each parameter we used the average and standard deviation of the 40 values as the amplitude and uncertainty ($\overline{a_{\mathrm{Xcos}}}\pm\delta a_{\mathrm{Xcos}}$, etc.). Fig. \ref{fig:GaussRayleighStats} shows that these are consistent with zero-mean Gaussians with identical widths. We then marginalized over the uninteresting phases of the $\tilde{\bm{E}}$ fields by combining the quadrature amplitudes to obtain total amplitudes, $a_{\mathrm{X}} = \sqrt{\abs{\overline{a_{\mathrm{Xcos}}}}^2 + \abs{\overline{a_{\mathrm{Xsin}}}}^2}$, etc., and total uncertainties, $\delta a_{\rm X} = (\delta a_{\mathrm{Xcos}} + \delta a_{\mathrm{Xsin}})/2$, etc. As expected these are consistent with Rayleigh distributions.

\begin{figure}[h!]
\centering \includegraphics[width=\columnwidth]{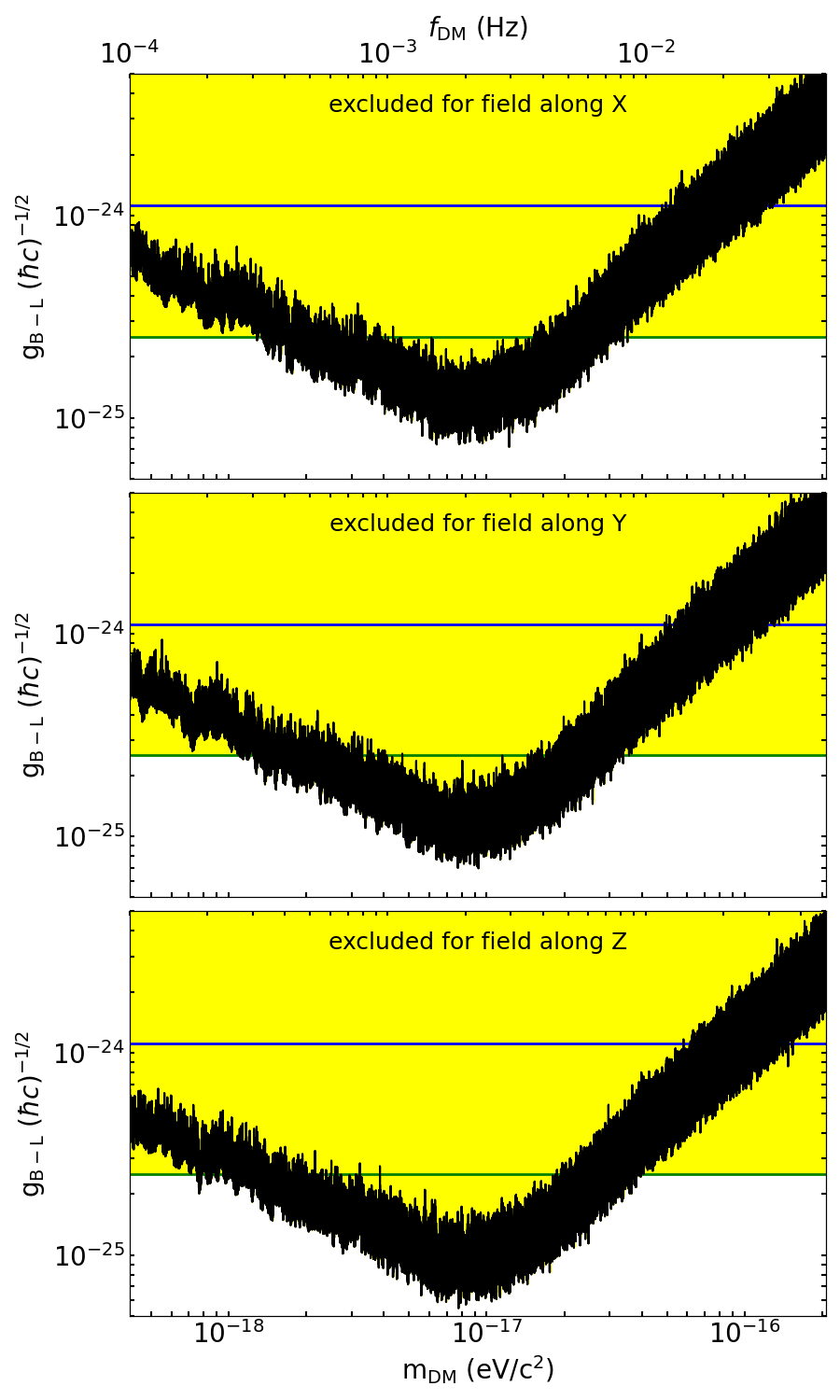}
\caption{95\% confidence upper limits on B-L coupled dark matter ULMBs. The exclusion regions are indicated in yellow. The black lines, which consist of 491,019 95\% confidence upper limits evenly spaced in frequency, show the limits on the amplitudes of fields along X, Y, and Z. (For reference, $g_{\rm B-L} /\sqrt{(\hbar c)} = 10^{-25}$ corresponds to a torque $9.73\times 10^{-19}$ N m). The horizontal blue lines are the upper limit from the E\"ot-Wash EP test Ref.~\cite{CQG2012}. The green lines are the upper limits from the initial MICROSCOPE result \cite{MICROSCOPE, MICROSCOPE2}.}
\label{fig:B-Llimits}
\end{figure}

\section{Results}
We calculated confidence limits, shown in Fig. \ref{fig:B-Llimits}, on waves propagating along $\hat{\bm{X}}$, $\hat{\bm{Y}}$, and $\hat{\bm{Z}}$ separately using the Rice distribution -- the generalization of the conventional mean$+ 2\sigma$ 95\% C.L. upper limit on normally distributed results to Rayleigh distributed total amplitudes appropriate for our signal search (computed with the percent-point function of the Rice distribution in \texttt{SciPy} \cite{scipy}). Note that below the free resonance of the torsion oscillator, the limits scale as $1/\sqrt{f}$ (behavior characterized, but not limited to internal damping). Above the resonance where the oscillator response drops as $1/f^2$, the limits scale as $f$ (which is evidence that, in our frequency range, the autocollimator noise amplitude is proportional to $1/f$). Because our results were taken for spans $S<t_{\rm c}$, they do not necessarily reflect the long-term average strengths of $\tilde{\bm{E}}$ which would be revealed by data taken with spans $S>t_{\rm c}$. Corrections for this effect have been discussed for scalar bosons \cite{scalarstochastic, scalargradstatistics, scalarwavedetection} and axions \cite{axionstatistics}. However, to our knowledge, corrections for $\tilde{\bm{E}}$ of a vector boson are not available (note that chaotic light is not a sufficient analog as photons have only 2 polarization states while vector ULMBs have three). Furthermore Ref.~\cite{scalarstochastic} assumes an unrealistic isotropic, isothermal, fully-virialized halo model. Recent analyses of Gaia-2 data \cite{DMhurricane, Fattahi, FIRElight} suggest that 10\%-50\% of the dark matter in our galaxy is in unequilibrated debris streams from recent mergers with satellite galaxies. As a result we cite an experimental result that is easily corrected later.

A viable dark matter candidate cannot consist of ULMBs with masses and couplings already excluded by conventional EP tests that provide strong constraints on the couplings of ULMBs over a large range of masses $m_{\rm DM} < \hbar c/R_{\rm Earth}=3.1\times 10^{-14}$ eV/$c^2$. If the dark matter predominantly consists of B-L coupled ULMBs, the constraints set by this work improve on MICROSCOPE's initial EP limits \cite{MICROSCOPE, MICROSCOPE2} in the mass range between $2 \times 10^{-18}$ eV/$c^2$ and $20 \times 10^{-18}$ eV/$c^2$ and improve on E\"ot-Wash EP limits \cite{CQG2012} in a slightly wider band. Conventional EP tests provide strong constraints on dark matter candidates, but as pointed out by Ref.~\cite{gr:13}, the signal in those tests is $\propto g_{\rm B-L}^2$, whereas in this search the signal is linear in $g_{B-L}$. Hence, an N fold improvement in the sensitivity would require an N$^2$ improvement in conventional EP results for an equal improvement in constraints.

\section{Conclusion}
We described a torsion-balance search for a dark matter signal from B-L coupled ULMB candidates. In future experiments we expect improved results using a longer and thinner fused-silica torsion fiber and a larger B-L dipole moment. Upgrades to the autocollimator and a better seismic environment will eliminate the need for a gravitational damper and allow for longer cuts, which would improve on the uncertainties of the fit parameters. Active tilt feedback could address the need to account for daily drifts allowing us to extend constraints to lower masses. 
We hope to improve our sensitivity by a factor of five or more.

We thank W.\,A.~Terrano for encouraging us to do this experiment. J.\,G.~Lee helped with our analysis methods. We thank Peter Graham and Will Terrano for clarifying conversations. This work was supported in part by National Science Foundation Grants PHY-1912514, PHY-2011520, and PHY-2012350.

\bibliographystyle{unsrtnat}
\bibliography{darkep-submit}

\clearpage
\section{Appendix}
\subsection{Constraints on ULMBs coupled to B and L}

\begin{figure}[H]
\centering \includegraphics[width=\columnwidth]{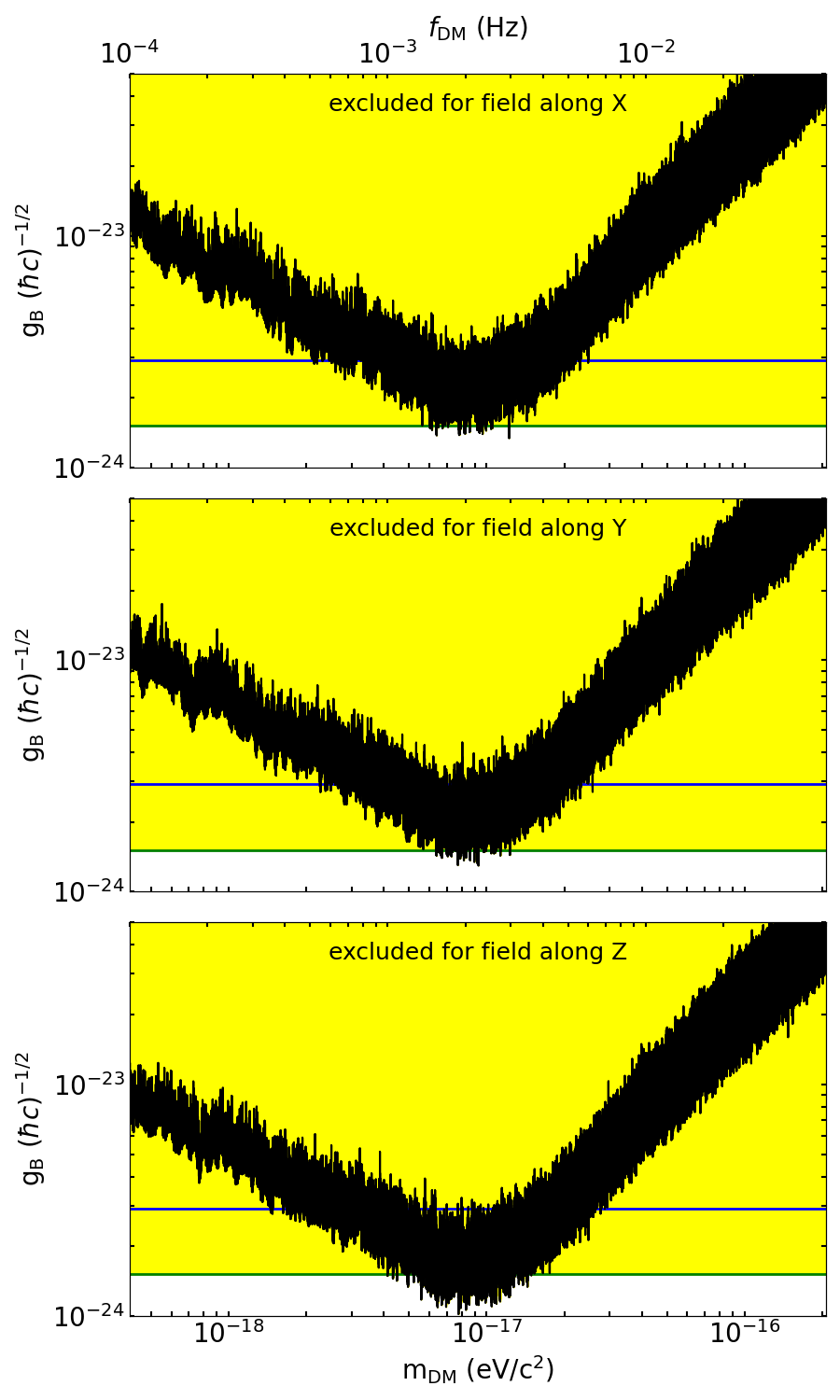}
\caption{Plot equivalent to Fig. \ref{fig:B-Llimits} for B-coupled dark matter ULMBs.}	
\label{fig:XYZlimits}
\end{figure}

We assumed above that the ULMBs are coupled to B-L. Our constraints on ULMBs coupled to B are shown in Fig. \ref{fig:XYZlimits}. Because $\Delta_{\rm B} \ll \Delta_{\rm B-L}$, our constraints on ULMBs coupled to L are similar to those shown in Fig. \ref{fig:B-Llimits}. In a narrow band of masses around our most sensitive mass $m_{\rm DM}=8\times 10^{-18}$ eV/$c^2$ there is some improvement on the E\"ot-Wash EP B limits. MICROSCOPE's result constrains this parameter space.\\
\newpage

\subsection{Analysis systematics}
\begin{figure}[H]
\centering \includegraphics[width=\columnwidth]{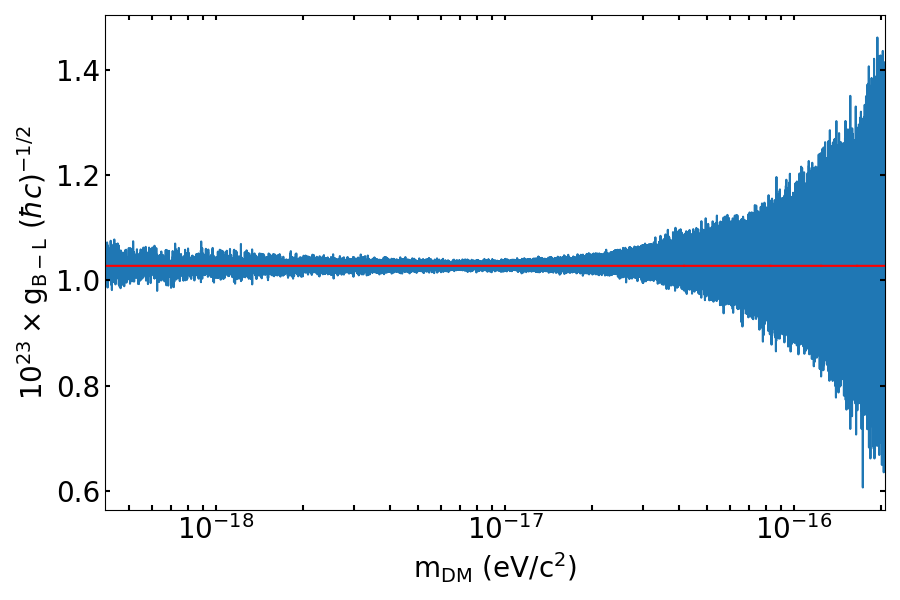}
\caption{A plot of X fit amplitudes resulting from an injected X signal with $g_{\rm B-L}(\hbar c)^{-1/2} = 10^{-23}$ (red line). The results are consistent with the expected amplitude indicating that the instrumental drift parameters did not appreciably bias our result.}
\label{fig:FullInjection}
\end{figure}

\begin{figure*}[!h]
\centering \includegraphics[width=\textwidth]{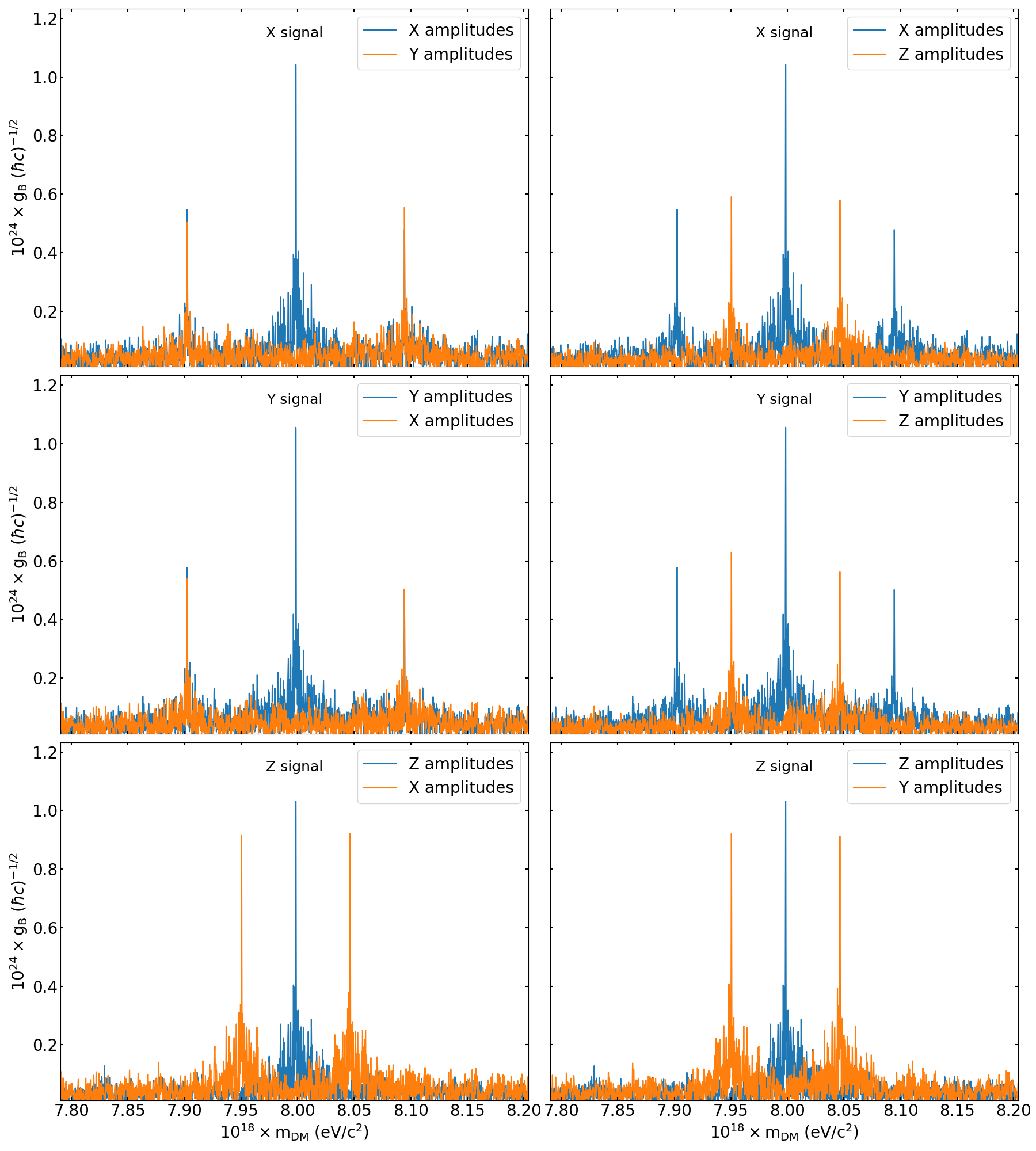}
\caption{$a_{\rm X}$, $a_{\rm Y}$, and $a_{\rm Z}$ amplitudes resulting from  injected X, Y, and Z signals with $g_{\rm B-L}(\hbar c)^{-1/2} = 10^{-24}$ and mass $8 \times 10^{-18}$ eV/$c^2$ ($f_{\mathrm{DM}}$ near the resonant frequency 1.934 mHz), where the frequency-dependent correction of Equation \eqref{torque_cor} is largest. Although X, Y, and Z signals produced peaks in all three amplitudes, the most prominent was always associated with the injected signal. The agreement between the input signal and the output peak amplitudes indicates no bias from the frequency-dependent correction or instrumental parameters.}	
\label{fig:XYZInjection}
\end{figure*}

\begin{figure*}[!h]
\centering \includegraphics[width=\textwidth]{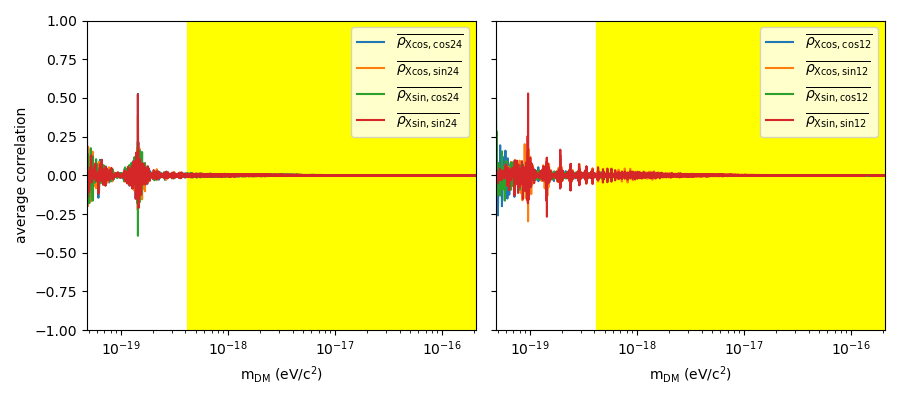}
\caption{Pearson correlation coefficients $\rho_{i,j}=cov(i,j)/(\sigma(i)\sigma(j))$ for the X quadrature basis functions and the 12 hour (Left) and 24 hour (Right) instrumental basis functions summed over all 3,334,034 data points at each mass. The plot for Y quadrature amplitudes is essentially the same. Z is less interesting because it has no sidereal modulation. The yellow region shows the masses included in the analysis. Starting at $0.4\times 10^{-18}$ eV/$c^2$ the correlation is small enough to get accurate fits (at worse $\pm$5\%) as evidenced by injection tests (Fig. \ref{fig:FullInjection} and \ref{fig:XYZInjection}).}
\label{fig:CovarDaily}
\end{figure*}

We did a variety of tests to verify the accuracy of our analysis. First, we predicted the angle $\theta$ in response to a torque signal along $\hat{\bm{X}}$ and added it to the angle data. We ran the entire analysis chain to obtain a resulting amplitude. As shown in Fig. \ref{fig:FullInjection}, it was consistent with the injected torque amplitude. Fig. \ref{fig:XYZInjection} shows the response to an injected signal near the resonant frequency of the balance where the frequency-dependent correction outlined in Equation \eqref{torque_cor} is largest. The most prominent peaks are consistent with the injected signal and demonstrate that the direction of a ULMB signal can be determined. The less prominent peaks occur because of covariance, for example, between $K_{\rm Xp}(t)e^{i\omega_{\rm DM}t}$ and $K_{\rm Yp}(t)e^{i(\omega_{\rm DM}\pm 2\omega_{\oplus})t}$ and $K_{\rm Zp}(t)e^{i(\omega_{\rm DM}\pm\omega_{\oplus})t}$ where $\omega_{\oplus}=7.3\times 10^{-5}$ rad/s is the sidereal frequency. Since $K_{\rm Zp}(t)$ is approximately constant, the less prominant peaks are spaced by $\omega_{\oplus}$. Previous results with a rotating torsion balance \cite{SPIN} mitigated this by taking data with multiple orientations of the dipole. Fig. \ref{fig:CovarDaily} shows that the covariance between the science and the daily instrumental basis functions, which could have introduced errors, is negligible.

\end{document}